\documentclass[aps,twocolumn,pra,showpacs,floatfix]{revtex4}
\usepackage{epsfig}
\usepackage{graphicx}
\usepackage{dcolumn}
\usepackage{amssymb,amsmath}
\usepackage{mathrsfs}
\begin{document}

\title{Shift of nuclear clock transition frequency in $^{229}$Th ions due to hyperfine interaction.}

\author{V. A. Dzuba and V. V. Flambaum}

\affiliation{School of Physics, University of New South Wales, Sydney 2052, Australia}

\begin{abstract}
We calculate hyperfine structure of $^{229}$Th and its ions (Th~IV, Th~III, Th~II, Th~I) to reveal the dependence of the nuclear clock frequency on the hyperfine interaction (hfi). We calculate first and second-order hfi shifts and demonstrate that due to the differences in the hyperfine structure for different ions and for the ground and isomeric nuclear states, the nuclear frequencies in Th~IV, Th~III, Th~II and Th~I are also different.
The first-order shift of frequency is large for a particular hyperfine component, but it vanishes after averaging over all hyperfine states. The second-order shift is small but it does not vanish after averaging. It is three to six orders of magnitude smaller than the shift of the nuclear frequencies due to the Coulomb  electron-nucleus  interaction considered in our previous work (V. A. Dzuba and V. V. Flambaum, arXiv:2309.11176 (2023)). 
However, it is six to eight orders of magnitude larger than the projected  accuracy of the nuclear clock ($10^{-19}$).

\end{abstract}

\maketitle

\section{Introduction}

The $^{229}$Th isotope is considered for building nuclear clocks of exceptional accuracy (see, e.g. ~\cite{PeikTamm,Thm}).
This is because the $^{229}$Th nucleus has a unique feature of having very low-energy excitation connected to the nuclear ground state by the magnetic dipole (M1) transition (see, e.g. reviews ~\cite{Rev,Rev1} and references therein). The relative uncertainty of the proposed clock is expected to reach $10^{-19}$~\cite{Th3+}. There are strong arguments that this nuclear clock would be very sensitive to physics beyond standard model including space-time variation of the fundamental constants, violation of the Lorentz invariance and Einstein equivalence principle, and search  for scalar and axion dark matter fields ~\cite{vara,Lorentz,varq,vara1,vara2,vara3,vara4,Arvanitaki,Stadnik}. 
There is good progress in recent years of measurements of the frequency of this transition~\cite{wn1,wn2,wn3,wn4,wn5}.
The latest, most precise measurements, give the value of 8.338(24)~eV~\cite{wn}.
There are plans to use Th ions of different ionisation degree~\cite{Th3+,Th+,dr} and solid-state Th nuclear clock~\cite{Hudson,ThSS,ThSS1}.
In our previous paper~\cite{previous} we demonstrated that in all these systems the frequency of the nuclear clock will be different. This is due to the Coulomb interaction of atomic electrons with the nucleus, leading to the significant electronic shift of the nuclear transition frequency.
The effect is similar to the field isotopic shifts  and isomeric shifts of atomic transition frequencies  but  here electrons  affect the nuclear transition. The shift  is caused by the difference in electronic structure between Th ions and the difference in the  nuclear radii for the ground and isomeric nuclear states.

In the present paper we consider another effect which leads to the difference in nuclear frequency for different Th systems.
It comes from the hyperfine interaction (hfi) of atomic electrons with the nucleus. The difference in frequencies is caused by the difference in electronic structure of Th ions and the difference in nuclear parameters (nuclear spin $I$, nuclear magnetic moment $\mu$ and nuclear electric quadrupole moment $Q$) for the ground and isomeric nuclear states.

The hyperfine structure (hfs) of Th and its ions has beed studied for many years both experimentally and theoretically~\cite{Th+old,ThQ,Th-mu-Q,Beloy,Th+hfs,ThIIIhfs}.
However, the data is still incomplete. There are no experimental data for the hfs of the ground states of Th~I and Th~III.
To the best of our knowledge, there are no calculations for Th~II and Th~I hfs. We address these shortcoming by performing the hfs calculations for Th~IV, Th~III, Th~II and Th~I. For the purpose of the present work we need only hfs of the ground states of all these systems. 
However, we calculate hfs constants for some excited states too for the convenience of comparing with other data either experimental or theoretical. This gives us some understanding of the accuracy of our calculations.

First-order hfi leads only to splitting of atomic levels. The total shift, averaged over hfs components, is zero. 
In contrast, the second-order hfi does lead to the $F$-dependent shifts of the levels which do not vanish after averaging. Here $F$ is the value of the total atomic angular momentum, which is the sum of electronic angular momentum  $J$ and nuclear spin  $I$ ($\mathbf{F}=\mathbf{J}+\mathbf{I}$).
In the  present paper we calculate both, the first-order hfs and the second-order hfi shift.

\section{First-order hfs shift}
First-order hfs leads to splitting of the atomic levels. Total shift averaged over hfs components with statistical weight $(2F+1)$ is zero.
However, it was suggested in Ref.~\cite{Th3+} to use the stretched hyperfine states ($F=I+J$, $F_z=F$ and $F_z=-F$) because in such states atomic wave function is a simple product of electronic and nuclear wave functions.   This absence of entanglement  between electron and nuclear variables leads to strong suppression of systematic effects  in the nuclear  transition since external fields mainly interact with electrons.  
 For a stretch state the energy shift caused by the magnetic dipole hyperfine interaction is
\begin{equation}\label{e:dea}
\Delta E_A = AIJ,
\end{equation}
where $A$ is the magnetic dipole hfs constant, $I$ is the nuclear spin, $J$ is the total electron angular momentum.
The difference in shifts for the ground and isomeric nuclear states comes from different values of nuclear magnetic moments $\mu$ of the states.
It is convenient to separate electronic and nuclear variables by writing $A=A_0 \mu/I$. Then the difference is given by
\begin{equation}\label{e:ddea}
\delta (\Delta E_A) = A_0J\Delta \mu.
\end{equation}
For the electric quadrupole hyperfine shift the general formula reads
\begin{eqnarray}\label{e:ddeb}
&&\Delta E_B = \frac{B}{2} \left[ C(C+1) -\frac{4}{3}J(J+1)I(I+1)\right] ,\\
&&C=F(F+1)-J(J+1)-I(I+1). \nonumber
\end{eqnarray}
To get the result for the stretched state we should  substitute $F=I+J$. This gives 
\begin{equation}\label{e:ddebstretch}
\Delta E_B = \frac{B}{3} I J (2I-1)(2J-1).
\end{equation} 
Separating electronic and nuclear variables by introducing $B=B_0Q$, where $Q$ is nuclear electric quadrupole moment, we see that the difference between ground and isomeric nuclear states comes from the different values of $Q$ and $I$. 
To calculate frequency differences caused by hfi we need to know electronic structure factors $A_0$, $B_0$ for the ground states of Th~IV, Th~III, Th~II, and Th~I, as well as nuclear parameters $\mu$, $I$ and $Q$ for the ground and isomeric nuclear states of $^{229}$Th.
Nuclear parameters are known~\cite{Thm,Th-mu-Q,ThQold}, while experimental data on the hfs of the ground states are available for Th~IV~\cite{ThQ,Th-mu-Q} and Th~II~\cite{Th+old,Th+hfs}. 

\begin{table} 
  \caption{\label{t:nuc}Nuclear parameters of the ground and isomeric states of $^{229}$Th taken from Refs.~\cite{Thm,Th-mu-Q,ThQold}.}
\begin{ruledtabular}
\begin{tabular}   {l ccc}
\multicolumn{1}{c}{Nucleus}&
\multicolumn{1}{c}{$I$}&
\multicolumn{1}{c}{$\mu$ [$\mu_N$]}&
\multicolumn{1}{c}{$Q$ [barn]}\\
\hline
$^{229}$Th    & 5/2  & 0.360(7)\footnotemark[1] & 3.11(6)\footnotemark[1];  3.15(3)\footnotemark[2] \\
$^{229m}$Th & 3/2   & -0.37(6)\footnotemark[3] &  1.73(10)\footnotemark[3] \\
\end{tabular}			
\end{ruledtabular}
\footnotetext[1]{Ref. \cite{Th-mu-Q}.}
\footnotetext[2]{Ref. \cite{ThQold}.}
\footnotetext[3]{Ref. \cite{Thm}.}
\end{table}

\subsection{Hyperfine structure of Th~IV}

\begin{table} 
  \caption{\label{t:AB}Experimental and theoretical values of the magnetic dipole and electric quadrupole hfs constants $A$ and $B$ (in MHz) for four low states of $^{229}$Th~IV. It is assumes in the calculations that $\mu=0.360\mu_N$, $I=5/2$~\cite{Thm}, and $Q=3.11b$~\cite{Th-mu-Q}.}
\begin{ruledtabular}
\begin{tabular}   {l cccc cccc}
\multicolumn{1}{c}{State}&
\multicolumn{1}{c}{$A_{\rm expt}$}&
\multicolumn{3}{c}{$A_{\rm theor}$}&
\multicolumn{1}{c}{$B_{\rm expt}$}&
\multicolumn{3}{c}{$B_{\rm theor}$}\\
&\multicolumn{1}{c}{\cite{ThQ}}&
\multicolumn{1}{c}{\cite{Th-mu-Q}}&
\multicolumn{1}{c}{\cite{Beloy}}&
\multicolumn{1}{c}{This}&
\multicolumn{1}{c}{\cite{ThQ}}&
\multicolumn{1}{c}{\cite{Th-mu-Q}}&
\multicolumn{1}{c}{\cite{Beloy}}&
\multicolumn{1}{c}{This}\\
&&&&\multicolumn{1}{c}{work}&&&&
\multicolumn{1}{c}{work}\\
\hline
$5f_{5/2}$ &  82.2(6) & 82.5  & 91.9  & 81.0 & 2269(6)  & 2254 & 2258 & 2264 \\
$5f_{7/2}$ &  31.4(7) & 31.0   & 36.6  & 31.0 & 2550(12) & 2515 & 2630 & 2569 \\
$6d_{3/2}$ & 155.3(12)& 155.3 &       &160  & 2265(9)  & 2295 & & 2276 \\
$6d_{5/2}$ & -12.6(7) & -13.2   &       & -27  & 2694(7)  & 2716 & & 2724 \\ 
\end{tabular}			
\end{ruledtabular}
\end{table}

Hyperfine structure of the ground state of the Th~IV is known experimentally~\cite{ThQ}. However, it is instructive to perform the calculations to check the accuracy of our approach. 

We start the calculations from the relativistic Hartree-Fock (RHF) procedure for the closed-shell Rn-like core. The single-electron basis states for valence space are calculated in the field of frozen core using the B-spline technique~\cite{B-spline}. These basis states are used to calculate the single-electron correlation operator $\hat \Sigma^{(2)}_1$ (correlation potential~\cite{CPM}). The same basis is used for the configuration interaction (CI) calculations in next subsection. We use the second-order many-body perturbation theory to calculate $\hat \Sigma^{(2)}_1$. 

The single-electron valence states of Th~IV are calculated using this correlation potential $\hat \Sigma^{(2)}_1$,
\begin{equation}\label{e:BO}
(\hat H^{\rm RHF} + \lambda \hat \Sigma^{(2)}_1 - \epsilon_v)\psi^{\rm BO}_v =0.
\end{equation} 
Here $\hat H^{\rm RHF}$ is the RHF Hamiltonian, index $v$ numerates valence states, $\lambda$ is rescaling coefficient, its value is chosen to fit experimental energies.
We use $\lambda=0.75$ for $d$-states and $\lambda=0.8$ for $f$-states. The fitting with $\lambda$ imitates the effect of higher-order correlations.
The solutions of (\ref{e:BO}) are usually called Brueckner orbitals (BO).

To include the interaction of atomic electrons with the magnetic dipole and electric quadrupole fields of the nucleus (i.e., to calculate hfs) we use the version of the random phase approximation (RPA) developed in Ref.~\cite{CPM}.
The RPA equations have the form
\begin{equation}\label{e:RPA}
(\hat H^{\rm RHF} - \epsilon_c)\delta\psi_c = -(\hat F + \delta V^F_{\rm core})\psi_c.
\end{equation} 
Here $F$ is the operator of external field (magnetic dipole or electric quadrupole field of the nucleus), index $c$ numerates core states, $\delta\psi_c$ is the correction to core orbital $\psi_c$ caused by external field, $\delta V^F_{\rm core}$ is the correction to the self-consistent RHF potential of the core caused by the change of all core orbitals. After the RPA equations are solved, the energy shift for valence state $v$ is given by
\begin{equation}\label{e:ME}
\delta \epsilon_v = \langle \psi^{\rm BO}_v|\hat F + \delta V^F_{\rm core} + \delta \Sigma^{(2)}_1|\psi^{\rm BO}_v\rangle.
\end{equation}
Here $ \delta \Sigma^{(2)}_1$ is the change of the correlation potential due to the change of all basis states in the external field.
Corresponding term is called {\em structure radiation}. It was not included in our previous calculations for the electric quadrupole hfs constant $B$~\cite{vara2}. 
Inclusion of the structure radiation leads to significant improvement of the results for both hfs constants $A$ and $B$.

Energy shifts (\ref{e:ME}) are used to calculate hfs constants $A$ and $B$. 
The results are presented in Table~\ref{t:AB}. Comparison with experimental data from Ref.~\cite{ThQ} and earlier calculations of Ref.~\cite{Th-mu-Q,Beloy} shows good agreement in all cases except $A$ constant for the $6d_{5/2}$ state. Here the value of $A$ is small and  dominated by the many-body corrections  which change the sign of $A$ .  

For the purpose of the present work we need only hfs constants $A$ and $B$ for the ground state. We can use experimental values for Th~IV.
However, the calculations are useful since they serve as a starting point for more complicated calculations for Th~III, Th~II and Th~I (see next subsection) where experimental data are incomplete.

\subsection{Hyperfine structure of Th~I, Th~II and Th~III.}

\begin{table*} 
  \caption{\label{t:th-hfs}Experimental and theoretical values of the magnetic dipole and electric quadrupole hfs constants $A$ and $B$ (in MHz) for some states of $^{229}$Th~III, $^{229}$Th~II, and $^{229}$Th~I. It is assumes in calculations that $\mu=0.360\mu_N$, $I=5/2$~\cite{Thm}, and $Q=3.11b$~\cite{Th-mu-Q}.}
\begin{ruledtabular}
\begin{tabular}   {llc r  cccc cccc}
\multicolumn{1}{c}{Ion/}&
\multicolumn{2}{c}{State}&
\multicolumn{1}{c}{Energy}&
\multicolumn{4}{c}{$A$ [MHz]}&
\multicolumn{4}{c}{$B$ [MHz]}\\
\multicolumn{1}{c}{Atom}&
\multicolumn{1}{c}{Conf.}&
\multicolumn{1}{c}{$J^p$}&
\multicolumn{1}{c}{Ref.~\cite{NIST}}&
\multicolumn{1}{c}{CI+MBPT}&
\multicolumn{1}{c}{MCDF}&
\multicolumn{1}{c}{This}&
\multicolumn{1}{c}{Expt.}&
\multicolumn{1}{c}{CI+MBPT}&
\multicolumn{1}{c}{MCDF}&
\multicolumn{1}{c}{This}&
\multicolumn{1}{c}{Expt.}\\

&&&&\multicolumn{1}{c}{Ref.~\cite{ThIIIhfs}}&
\multicolumn{1}{c}{Ref.~\cite{ThIIIhfs}}&
\multicolumn{1}{c}{work}&
\multicolumn{1}{c}{Ref.~\cite{Thm}}&
\multicolumn{1}{c}{Ref.~\cite{ThIIIhfs}}&
\multicolumn{1}{c}{Ref.~\cite{ThIIIhfs}}&
\multicolumn{1}{c}{work}&
\multicolumn{1}{c}{Ref.~\cite{Thm}}\\
\hline
Th~III & $5f6d$   & $4^-$  &     0   &  64(17) &  81(4)   & 76.3 &           & 3287(630)& 3008(260) & 3828 &        \\
           & $6d^2$  & $2^+$  &    63  & 143(47) & 162(8) & 146  & 151(8)&      68(23)&       71(7)   & -35 & 73(27) \\
           & $5f^2$   & $4^+$  & 15148 &  38(3)  &  72(3) & 60.2 &             & 1221(390)& 1910(200) & 2469&         \\
           & $5f6d$  & $1^-$  & 20711 & 109(36) &  90(4) & 111  &  88(5)&  839(220)&  689(110) & 672 & 901(18) \\
           & $5f^2$  & $4^+$  & 21784 &   8(36)   &  26(2) & 4.9  &       &    65(21)&    39(45) & -600&         \\
Th~II   & $6d^7s$&$3/2^+$&        0  &              &          & -472 & -444.2(3.4)\footnotemark[1]  &  &  & 507 & 308(13)\footnotemark[1] \\
           &              &              &           &              &          &          & -444.2(1.9)\footnotemark[2]  &  &  &       & 303(6)\footnotemark[2] \\
Th~I    & $6d^27s^2$&$2^+$&        0  &              &          &  41 & & & & 320 & \\
\end{tabular}
\footnotetext[1]{Ref.~\cite{Th+hfs}.} 			
\footnotetext[2]{Ref.~\cite{Th+old}.} 			
\end{ruledtabular}
\end{table*}

To calculate hfs of atoms with two, three and four valence electrons we use the all-orders SD+CI~\cite{SD+CI} ( single-double coupled cluster method combined with the configuration interaction technique). The SD method produces the all-orders single electron correlation operator $\Sigma^{\infty}_1$, similar to the second-order operator $\Sigma^{(2)}_1$ used in the  previous section. It also produces the all-orders  two-electron correlation operator $\Sigma^{\infty}_2$. The use of the all-orders correlation operators usually leads to more accurate results. The effective CI Hamiltonian has the form
\begin{equation}\label{e:CI}
\hat H^{\rm CI} = \sum_{i=1}^{N_v} \left(\hat H^{\rm RHF} + \hat \Sigma^{\infty}_1\right)_i + \sum_{i<j}\left(\frac{e^2}{r_{ij}} + \hat \Sigma^{\infty}_{2ij}\right).
\end{equation}
Here summation goes over valence electrons, $N_v$ is the number of valence electrons (2,3 or 4).
The same RPA technique as in previous section is used to include external field. The energy shift for valence state $v$ is given by
\begin{equation}\label{e:RPA-ME}
\delta E_v = \langle v |\sum_i^{N_v}(\hat F+\delta V^F_{\rm core})_i|v\rangle.
\end{equation}
The results of calculations are presented in Table~\ref{t:th-hfs} and compared with available experimental data and previous calculations. 
Comparison shows that the uncertainty of the present calculations for the magnetic hfs constant $A$ is probably within 10\%. The uncertainty for $B$ is larger, it is about 30\% for large $B$ constants and even higher for small constants. There are no experimental data for the hfs of the ground states of Th~III and Th~I. There are also no other calculations for Th~I.

\subsection{Frequency shifts}

\begin{table} 
  \caption{\label{t:dab}Frequency shifts in the stretched  states caused by the first order hfi. The numbers in square brackets stand for powers of ten.}
\begin{ruledtabular}
\begin{tabular}   {lc ccc}
\multicolumn{1}{c}{Ion/}&
\multicolumn{1}{c}{$J_0$}&
\multicolumn{1}{c}{$\delta(\Delta E_A)$}&
\multicolumn{1}{c}{$\delta(\Delta E_B)$}&
\multicolumn{1}{c}{$\Delta \omega_N/\omega_N$} \\
\multicolumn{1}{c}{Atom}&&
\multicolumn{1}{c}{[MHz]}&
\multicolumn{1}{c}{[MHz]}& \\
\hline
   Th~IV & 5/2 & -1.049[+3] &   -4.667[+1] &   -5.436[-7] \\
   Th~III &  4  &-1.558[+3] &  -1.307[+2] &   -8.377[-7] \\
   Th~II  &  3/2 & 3.399[+3] &   -1.400[+1] &   1.679[-6] \\
   Th~I    &   2 &-4.145[+2] &   -2.800[+1] &   -2.195[-7] \\


\end{tabular}
\end{ruledtabular}
\end{table}

To calculate the first-order hfs shift between ground and excited nuclear states in the stretched atomic state we use formulae (\ref{e:ddea}) and (\ref{e:ddebstretch}), nuclear parameters from Table~\ref{t:nuc}, and the hyperfine structure constants $A$ and $B$ for the ground states of Th~IV, Th~III, Th~II, and Th~I  from Tables \ref{t:AB} and \ref{t:th-hfs}. For Th~IV and Th~I ions we use experimental values. For Th~III ion we use weighted theoretical values from Table 
\ref{t:th-hfs}. We assume theoretical uncertainty of 10\% for the hfs constant $A$  and 30\% for $B$  in  our results to use in  this weighting procedure. The results for the first order energy shifts in the stretched states are presented in Table~\ref{t:dab}.

\section{Second-oder hfs shift.}

\begin{table*} 
  \caption{\label{t:hfs2}Second-order hfs shift (MHz).
  See Eq.~(\ref{e:sum}) for definitions of $\Delta E_a$, $\Delta E_{ab}$, and $\Delta E_b$.
  $\langle \Delta E\rangle$ is the total shift averaged over all hfs components.
  The numbers in square brackets stand for powers of ten.}
\begin{ruledtabular}
\begin{tabular} {c cccccccc}
&\multicolumn{4}{c}{Ground nuclear state, $I=5/2$.}&
\multicolumn{4}{c}{Isomeric nuclear state, $I=3/2$.}\\
\multicolumn{1}{c}{$F$}&
\multicolumn{1}{c}{$\Delta E_a$}&
\multicolumn{1}{c}{$\Delta E_{ab}$}&
\multicolumn{1}{c}{$\Delta E_b$}&
\multicolumn{1}{c}{Sum}&
\multicolumn{1}{c}{$\Delta E_a$}&
\multicolumn{1}{c}{$\Delta E_{ab}$}&
\multicolumn{1}{c}{$\Delta E_b$}&
\multicolumn{1}{c}{Sum}\\
\hline
\multicolumn{9}{c}{Th~IV, $5f \ J_0=5/2$}\\
 1.0 &  2.87[-5] & -2.84[-4] &  7.01[-4] &  4.46[-4] & &&& \\
 2.0 &  7.76[-5] & -5.96[-4] &  1.15[-3] &  6.27[-4] &  1.04[-4] &  5.49[-4] &  7.24[-4] &  1.38[-3] \\
 3.0 &  1.29[-4] & -5.68[-4] &  6.24[-4] &  1.85[-4] &  1.95[-4] &  3.43[-4] &  1.51[-4] &  6.88[-4] \\
 4.0 &  1.58[-4] &           &           &  1.58[-4] &  1.95[-4] & -5.71[-4] &  4.19[-4] &  4.23[-5] \\
 5.0 &  1.29[-4] &  7.10[-4] &  9.74[-4] &  1.81[-3] & &&& \\
&\multicolumn{3}{c}{$\langle \Delta E\rangle$}&  7.54[-4] & \multicolumn{3}{c}{$\langle \Delta E\rangle$} & 5.03[-4] \\

\multicolumn{9}{c}{Th~III, $5f6d \ J_0=4$}\\
 


 1.5 & 1.24[-3] &  1.03[-3] &  9.77[-4] &  3.25[-3] &&&& \\
 2.5 & 1.76[-3] &  9.25[-4] &  2.90[-3] &  5.58[-3] &  1.68[-3] & -6.00[-4] &  2.15[-3] &  3.23[-3] \\
 3.5 & 2.33[-3] &  7.67[-4] &  5.55[-3] &  8.64[-3] &  2.96[-3] & -6.74[-4] &  3.90[-3] &  6.18[-3] \\
 4.5 & 2.76[-3] &  4.88[-4] &  5.52[-3] &  8.77[-3] &  3.71[-3] & -3.66[-4] &  1.21[-4] &  3.47[-3] \\
 5.5 & 2.84[-3] & -1.04[-4] &  1.42[-4] &  2.88[-3] &  3.30[-3] &  1.05[-3] &  1.08[-4] &  4.46[-3] \\
 6.5 & 2.31[-3] & -1.39[-3] &  2.70[-4] &  1.19[-3] &&&& \\

&\multicolumn{3}{c}{$\langle \Delta E\rangle$}&  4.72[-3] &\multicolumn{3}{c}{$\langle \Delta E\rangle$}& 4.36[-3] \\
\multicolumn{9}{c}{Th~II, $6d7s^2 \ J_0=3/2$}\\



 0.0 &           &           &           &           &  5.26[-2] & -1.11[-2] &  8.12[-4] &  4.23[-2] \\
 1.0 &  8.41[-2] &  1.03[-2] &  9.51[-4] &  9.53[-2] &  1.25[-1] & -3.31[-3] &  4.29[-4] &  1.22[-1] \\
 2.0 &  1.41[-1] &  1.56[-3] &  6.40[-4] &  1.44[-1] &  2.28[-1] & -1.32[-4] &  4.30[-4] &  2.28[-1] \\
 3.0 &  1.93[-1] &  3.74[-4] &  7.84[-4] &  1.94[-1] &  2.76[-1] &  3.10[-3] &  1.58[-4] &  2.80[-1] \\
 4.0 &  1.95[-1] & -4.59[-3] &  3.33[-4] &  1.91[-1] &&&& \\

&\multicolumn{3}{c}{$\langle \Delta E\rangle$}&  1.70[-1]  &\multicolumn{3}{c}{$\langle \Delta E\rangle$}& 2.19[-1] \\
\multicolumn{9}{c}{Th~I, $6d^27s^2 \ J_0=2$}\\

 0.5 &  4.42[-4] &  3.27[-4] &  1.52[-4] &  9.21[-4] &  4.91[-4] & -2.42[-4] &  2.11[-4] &  4.59[-4] \\  
 1.5 &  6.25[-4] &  4.23[-4] &  3.87[-4] &  1.44[-3] &  8.61[-4] & -3.42[-4] &  1.24[-3] &  1.75[-3] \\
 2.5 &  8.50[-4] &  3.40[-4] &  2.11[-3] &  3.30[-3] &  1.24[-3] & -9.87[-5] &  1.11[-4] &  1.25[-3] \\
 3.5 &  9.97[-4] & -4.45[-5] &  2.22[-4] &  1.17[-3] &  1.26[-3] &  3.06[-4] &  2.42[-5] &  1.59[-3] \\
 4.5 &  8.97[-4] & -4.03[-4] &  6.03[-5] &  5.54[-4] &&&& \\
&\multicolumn{3}{c}{$\langle \Delta E\rangle$}&  1.41[-3]  &\multicolumn{3}{c}{$\langle \Delta E\rangle$}& 1.41[-3] \\

\end{tabular}
\end{ruledtabular}
\end{table*}

\begin{table} 
  \caption{\label{t:ions}The difference in nuclear frequencies between Th ions due to the second-order hfs shifts.
  The average shifts $\langle \Delta E\rangle$ from Table~\ref{t:hfs2} are used to obtain the numbers.
  The numbers in square brackets stand for powers of ten.}
\begin{ruledtabular}
\begin{tabular} {lcl rrr }
\multicolumn{3}{c}{Ions}&
\multicolumn{2}{c}{$\Delta \omega_N$}&
\multicolumn{1}{c}{$\Delta \omega_N/\omega_N$}\\
&&&\multicolumn{1}{c}{(MHz)}&
\multicolumn{1}{c}{(eV)}& \\
\hline

Th I    &$-$& Th II  &  -4.916[-2] &   -2.033[-10] &   -2.438[-11] \\
Th I    &$-$& Th III &   3.495[-4] &    1.446[-12] &    1.734[-13] \\
Th I    &$-$& Th IV  &   2.459[-4] &    1.017[-12] &    1.220[-13] \\
Th II   &$-$& Th III &   4.951[-2] &    2.048[-10] &    2.456[-11] \\
Th II   &$-$& Th IV  &   4.940[-2] &    2.043[-10] &    2.450[-11] \\
Th III  &$-$& Th IV  &  -1.037[-4] &   -4.287[-13] &   -5.142[-14] \\
\end{tabular}
\end{ruledtabular}
\end{table}

The second-order hfs shift is given by
\begin{eqnarray}\label{e:hfs2} 
&&\Delta E_v = \sum_n \left[ \frac{\langle v|\hat H_A|n\rangle^2}{E_v-E_n} + \right. \\
&& \left. 2\frac{\langle v|\hat H_A|n\rangle \langle n|\hat H_B|v\rangle}{E_v-E_n} + \frac{\langle v|\hat H_B|n\rangle^2}{E_v-E_n}\right] \nonumber \\
&& \equiv \Delta E_a + \Delta E_{ab} + \Delta E_b. \label{e:sum}
\end{eqnarray}
Here $\hat H_A$ and $\hat H_B$ are the magnetic dipole and electric quadrupole hfi operators.
Corresponding reduced matrix elements are given by (see, e.g. Ref.~\cite{hfs2})
\begin{eqnarray}\label{e:hfsA}
&&\langle J_v,F|\hat H_A|J_n,F\rangle = (-1)^{I+F+J_n}\mu \times \\
&&\sqrt{\frac{(2I+1)(I+1)}{I}}\left\{\begin{array}{ccc} I & J_n & F \\ J_v & I & 1 \end{array} \right\} \langle J_v ||\hat H^e_A||J_n \rangle, \nonumber
\end{eqnarray}
and
\begin{eqnarray}\label{e:hfsB}
&&\langle J_v,F|\hat H_B|J_v,F\rangle = (-1)^{I+F+J_n}\frac{Q}{2} \times \\
&&\sqrt{\frac{(2I+1)(I+1)(2I+3)}{I(2I-1)}}\left\{\begin{array}{ccc} I & J_n & F \\ J_v & I & 2 \end{array} \right\} \langle J_v ||\hat H^e_B||J_n \rangle. \nonumber
\end{eqnarray}

We include only limited number of low-energy electron states in the summation over $n$ in (\ref{e:hfs2}).
They dominate over the rest of the sum due to small energy denominators.
Saturation of the summation has been observed on the level of few per cent. 

Nuclear excited states $| n\rangle$ also give a contribution to the sum in the expression for   $\Delta E_v$ in Eq. (\ref{e:hfs2}). However,  a minimal energy denominator $E_v-E_n$ in this case is the interval between the ground  and isomeric nuclear  states equal to 8.3 eV. This interval is 1-2 orders  of magnitude bigger than the interval between electron states (fine structure intervals in this case). As a result, the nuclear excitations contribution may be neglected in comparison with the electron excitations contribution. 

As in case of the first-order hfs shifts, the second-order shift (\ref{e:hfs2}) is different for the ground and isomeric nuclear states due to different nuclear parameters $\mu$, $I$ and $Q$. The results of calculations are presented in Tables~\ref{t:hfs2} and \ref{t:ions}. The second order hfi shift is much smaller than the first order shift.  
However,  the first-order shift vanishes after averaging over hfs components while the second-order shift does not.
Note that  the second-order hfi  shift is three to six orders of magnitude smaller than the shift due to the electrostatic interaction between atomic electrons and the nucleus  considered in our previous work~\cite{previous}. However, it is six to eight orders of magnitude larger than the projected relative uncertainty of the nuclear clock ($10^{-19}$).

\acknowledgments

This work was supported by the Australian Research Council Grants No. DP230101058 and DP200100150.

\end{document}